\documentclass[journal=jacsat,manuscript=article]{achemso}
\usepackage{graphicx}
\usepackage{bm}
\usepackage{amsmath}
\usepackage{amssymb}
\usepackage{hyperref}

\author{Wei Guo}
\affiliation{Frontier Institute of Science and Technology, and State Key Laboratory for Mechanical Behavior of Materials, Xi'an Jiaotong University, 710054, Xi'an, P. R. China.}

\author{Zhao Wang}
\email{wzzhao@yahoo.fr}
\affiliation{Frontier Institute of Science and Technology, and State Key Laboratory for Mechanical Behavior of Materials, Xi'an Jiaotong University, 710054, Xi'an, P. R. China.}

\author{Ju Li}
\email{liju@mit.edu}
\affiliation{Department of Nuclear Science and Engineering and Department of Materials Science and Engineering, Massachusetts Institute of Technology, Cambridge, Massachusetts 02139, USA}
\affiliation{Frontier Institute of Science and Technology, and State Key Laboratory for Mechanical Behavior of Materials, Xi'an Jiaotong University, 710054, Xi'an, P. R. China.}

\title{Diffusive versus displacive contact plasticity of nanoscale asperities: Temperature- and velocity-dependent strongest size}

\begin{document}
\begin{abstract}
We predict a strongest size for the contact strength when asperity radii of curvature decrease below ten nanometers. The reason for such strongest size is found to be correlated with the competition between the dislocation plasticity and surface diffusional plasticity. The essential role of temperature is calculated and illustrated in a comprehensive asperity size-strength-temperature map taking into account the effect of contact velocity. Such a map should be essential for various phenomena related to nanoscale contacts such as nanowire cold welding, self-assembly of nanoparticles and adhesive nano-pillar arrays, as well as the electrical, thermal and mechanical properties of macroscopic interfaces.\\
keywords: {material strength, dislocation plasticity, surface diffusion, sub-10nm, Zener-Hollomon scaling}
\end{abstract}

When two macroscopic solids touch, the atomistic realities of their
nanoscale contacts are hidden from easy view, but they actually
control how heat, electrical charge and forces are transferred across
the rough interface \cite{ferguson91science}. The true contact area
$A_{\rm true}$, defined by atoms of the two bodies that truly interact
atomistically (within certain interatomic force/distance cutoffs), is
usually much smaller than the nominal macroscopic contact area $A$.
$A_{\rm true}/A$ usually decreases with increasing surface roughness
of the two bodies, and increases with externally applied pressure
$P_{\rm ext}\equiv -F_{\rm ext}/A$.  Recently, Pastewka and Robbins
showed numerically using linear elasticity and half-space Green's
function how $A_{\rm true}/A$ depends on $P_{\rm ext}$ (e.g. linearly) for
two self-affine random surfaces, statistically self-similar within
profile wavelengths $[\lambda_s,\lambda_L]$ \cite{pastewka14pnas}.
They found that when the solids are elastically compliant enough, the
ratio between $A_{\rm true}/A$ and $P_{\rm ext}$ diverges due to
microscopic adhesion, signifying a ``non-sticky''-to-``sticky''
transition of the macro-contact.

While Pastewka and Robbins' results are revealing, the assumptions of
linear elasticity, especially at the lower wavelength cutoff
$\lambda_s$ ``of order nanometers''\cite{pastewka14pnas}, could be limiting. This is
because plasticity by dislocation motion and/or diffusion can occur,
certainly at high enough $P_{\rm ext}$, but may also occur at $P_{\rm
ext}=0$, as we show below.  One may also ask what could be a
physical basis for the $\lambda_s$ cutoff in solving elasticity
problems - is this assumed initial condition reflecting prior 
history, with surface diffusional plasticity\cite{Li15NM} that tends to smooth out
profile roughnesses finer than $\lambda_s$?  Incidentally, for
nanostructures, Jiang {\em et al.} \cite{jiang04}, and Guisbiers and
Buchaillot \cite{guisbiers08} have proposed size-dependent effective
diffusivity
\begin{equation}
 D(T,R) \;=\; D_{0\infty}\exp\left[ - \frac{CT_{\rm m\infty}}{k_{\rm
 B}T}\left(1-\frac{\tilde{\alpha}}{2R}\right) \right],
 \label{SizeDependentEffectiveDiffusivity}
\end{equation}
affecting nanoscale creep, that accompanies the well-established
melting-point reduction \cite{BuffatB76,ZhangESOKLWGA00}:
\begin{equation}
 T_{\rm m}(R) \;=\; T_{\rm m\infty}\left(1-\frac{\alpha}{2R}\right),
 \label{SizeDependentMeltingPoint}
\end{equation}
where $R$ is the radius of curvature of the nano-asperity, $T_{\rm
m\infty}$ is the bulk thermodynamic melting point, $k_{\rm B}$ is the
Boltzmann constant, and $C,\alpha,\tilde{\alpha},D_{0\infty}$ are
temperature- and size-independent positive constants. Such
``exponentially accelerated'' small-size diffusive kinetics
Eq.(\ref{SizeDependentEffectiveDiffusivity}) seem to have some
experimental support \cite{DickDZM02,ShibataBZMVG02}. While the
physical basis for Eq.(\ref{SizeDependentEffectiveDiffusivity}) is not
as well-understood as Eq.(\ref{SizeDependentMeltingPoint}), one notes
that in the $R<10$nm, and lower-homologous-temperature deformation
regime that we are mostly interested in, the effective diffusivity
$D(T,R)$ is dominated by the surface diffusion contribution. The
activation energy $Q_{\rm S}$ of surface diffusion mathematically
could have a leading-order correction proportional to $1/R$ in an
asymptotic expansion with respect to curvature, that physically could
be due to, for example, elasticity effect of the saddle-point configuration of diffusion, or
the ratio of atoms near surface
crystallographic facet-facet intersections (``surface defects'') among
all surface atoms. In other words, the curvature effect on surface
diffusion may be explained by the curvature-dependent concentration 
and mobility
of
``surface defects''. Surface diffusion could be the key for
understanding $\lambda_s$.  Recently, it was demonstrated
experimentally that under an external load, or a capillarity-generated
Young-Laplace pressure, plasticity by surface diffusion can indeed
happen at sub-10nm lengthscale at room temperature \cite{sun14nat,Xie15WangNM}.

\begin{figure}[thp]
\centerline{\includegraphics[width=13cm]{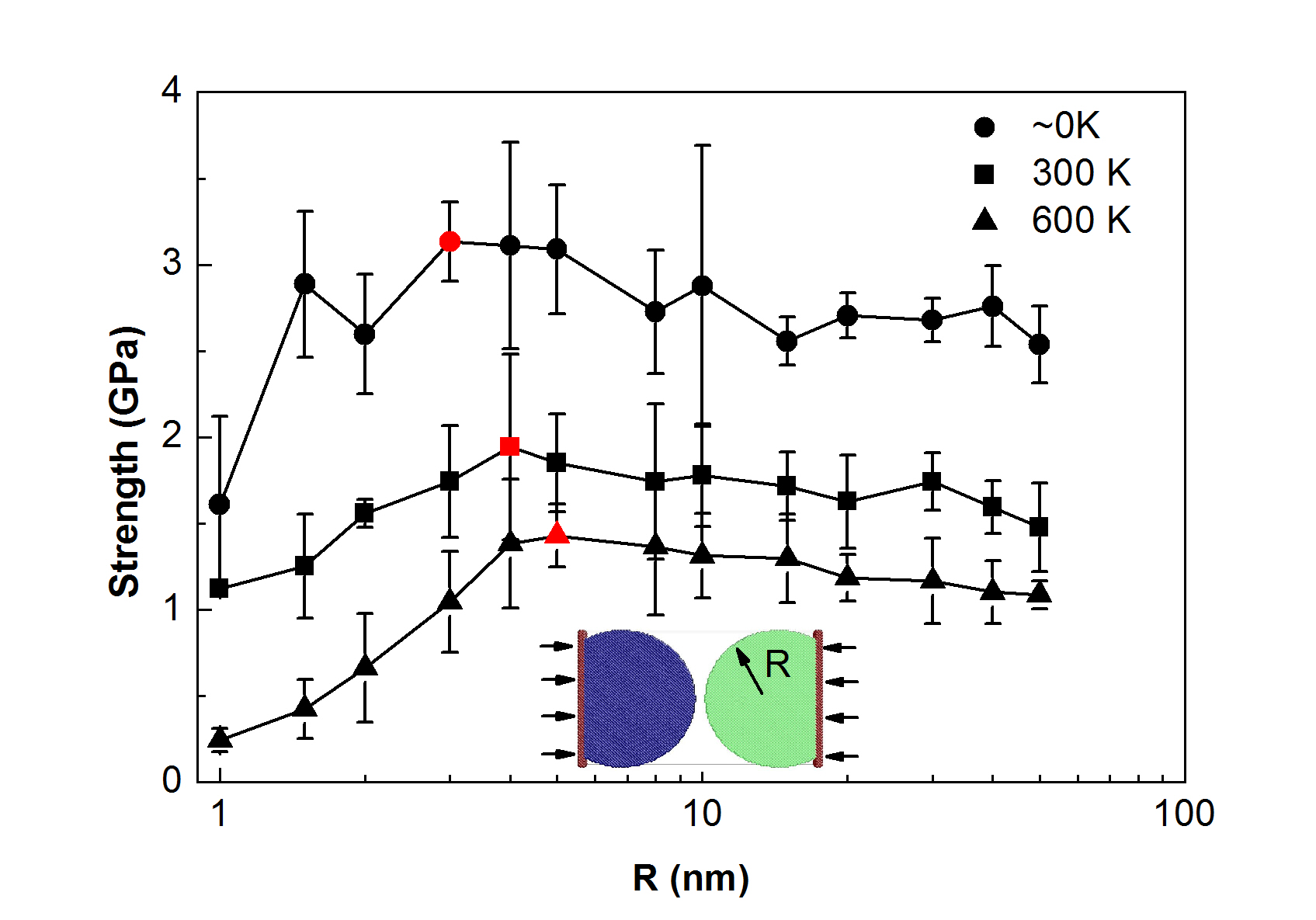}}
\caption{\label{fig:1} Contact strength as a function of the asperity radius $R$ at different temperatures. The peak values corresponding to the strongest size are marked with red symbols. Error bars are based on the deviation of difference in three crystal orientations. Note that the strength is averaged over three crystal orientations ([100] vs. [100], [100] vs. [110], and [100] vs. [16 5 0]) with respect to the two face-centered-cubic crystals.}
\end{figure}

With the above motivation, it is critical to understand the characteristics of plasticity for nanoscale asperities. Many experiments have shown that individual nano-structures can sustain close to their ideal strength \cite{ZhuL10} due to dislocation starvation. The ``smaller is stronger'' trend provides a strategy for increasing the material strength by nanostructuring. However when $R$ goes down to even smaller, surface diffusion could cause dramatic softening and ``smaller is much weaker'' \cite{tian13SR}. Here we look into this issue of diffusive versus displacive contact plasticity by atomistic simulations, using the classical molecular dynamics (MD) simulator LAMMPS \cite{Plimpton95}. As shown in the inset of Fig.\ref{fig:1}, in our simulations, two identical metal cylinders are moved towards each other. An embedded atom method potential \cite{zope03prb} was used to describe the atomistic interaction of Al,
which is chosen because of its elastic isotropy that simplifies the analysis. We have applied displacement control $-2\Delta x(t)$ between the two rigid outer boundaries in our simulations. To contrast the
outcome of different-size nano-asperities, we define total strain as
$\varepsilon \equiv \Delta x(t)/R$.  The stress is defined by the
engineering stress convention $\sigma_{\rm engineering} \equiv F /A$,
where $F$ is the computed total force sustained in one of the rigid
outer boundaries, and $A\equiv \pi(2R)L$ is the initial projected cross-sectional area of the cylinder, which is a ``nominal'' contact area in this simulation.  In this paper, the
``strength'' of contact is defined as the time-average of $\sigma_{\rm
engineering}$ in the strain range $0.08-0.2$ during loading, so it
should be interpreted as plastic ``flow strength'', and not the
initiation or yield strength (see Supporting Information). $R$ is varied from $1$ to $50$nm and the
strain rate $\dot{\varepsilon}$ is about $10^8$-$10^9$/s. MD simulations were
performed using a Nos$\acute{e}$-Hoover thermostat, and the systems
were relaxed for $2$ps at each load step of $0.1$\AA. We note that the strain rate range $10^8$-$10^9$/s in our simulations does not induce significant difference in the final contact strengths reported.

The size-strength relationship of our cylindrical contacts at $T=0$K,
300K and 600K are plotted in Fig.\ref{fig:1}. It clearly shows that
there exists a strongest size for the contact strength, below which
the ``smaller is stronger'' trend no longer holds. The sharp decrease
in strength when the contact size goes down to sub-10-nm scale
contradicts with the lattice dislocation-mediated deformation
mechanism \cite{greer05acta,zhu08prl}, and suggests that the strongest
size is in a deformation mechanism transition zone. In our simulated
samples, the critical sizes range from 5 to 10nm.  In this size range,
the difference between crystal and liquid surface energies causes the
crystal melting point $T_{m}$ to decrease as described by
Eq.(\ref{SizeDependentMeltingPoint}), and this effect is particularly
significant in the case of $R<$10nm \cite{ZhangESOKLWGA00}. Thus, it
can be expected that the surface atom diffusion may become important
for the strongest size, especially when the nano-asperity is under a high load.

\begin{figure}[thp]
\centerline{\includegraphics[width=13cm]{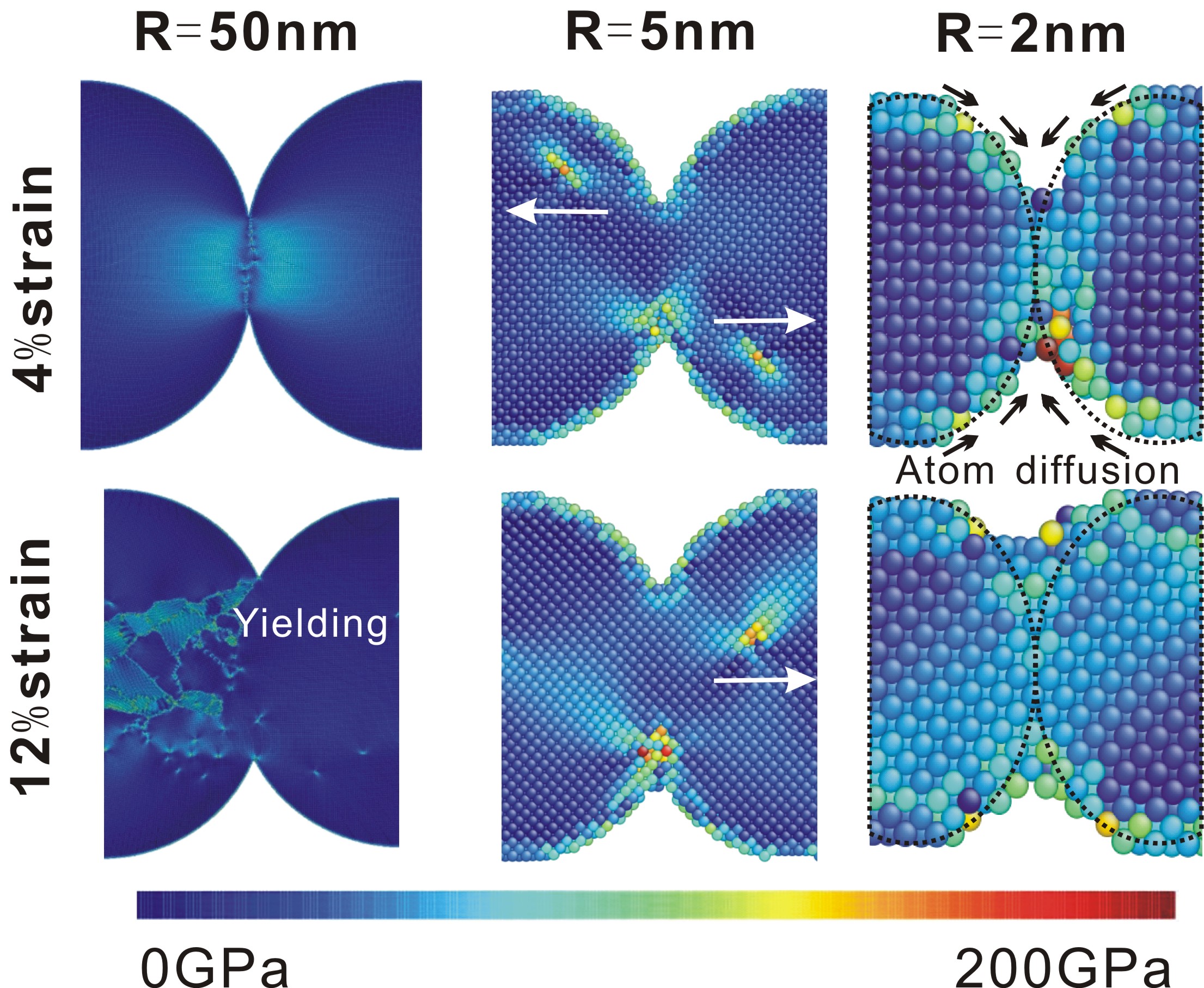}}
\caption{\label{fig:2} Snapshots of simulated room-temperature
contacts of three different sizes at two different applied strain
levels. The color of atoms represents the Von-Mises stress
distribution. The arrows in the middle panel show the direction of
dislocation displacement during loading. The arrows in the right panel
show the direction of atomistic diffusion flow.}
\end{figure}

We have performed detailed analyses of the atomistic configurations
and stress distributions in the simulated samples, as shown in
Fig.\ref{fig:2}. A classical Hertzian stress pattern is found beneath
the interface of the largest contact ($R=50$nm) at $4\%$ strain,
and the system started to deforms plastically before $12\%$ strain. In
a smaller contact ($R=5$nm), a pair of dislocations is found to form
beneath the interface, and move to interior following  increasing
compressive load. After the dislocation activity, a relaxed and more
homogenously distributed elastic stress results in the contacting
bodies. When $R$ decreases further down to $2$nm, we visually found that
the surface atoms diffuse significantly even at room
temperature, in agreement with first-principles
calculations \cite{sorensen96prl}. The atoms inside remain highly
crystalline while the surface atoms diffuse to the neck
region. Similar room temperature `liquid-like' deformation behaviors
were observed in experiments on Ag nanoparticles of about 10nm
diameter \cite{sun14nat}, as well as the cold welding of Au nanowires
($3$ to $10$nm diameter) \cite{lu10nat}.

\begin{figure}[htp]
\centerline{\includegraphics[width=13cm]{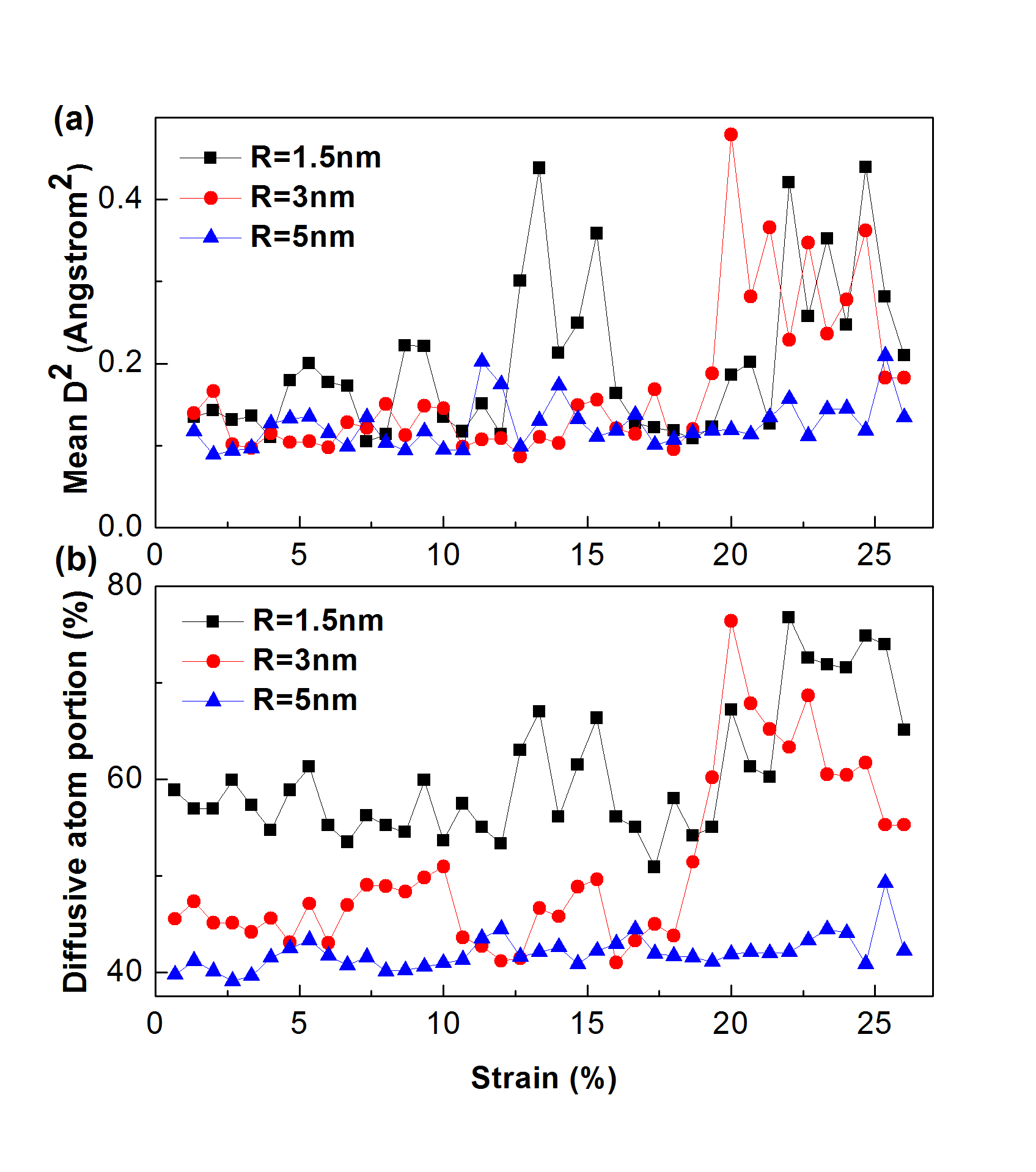}}
\caption{\label{fig:3} (a) Mean $D^2$ as a function of the strain in
three contacts of different curvature radii. (b) Diffusive atom ratios
as a function of the strain.}
\end{figure}

To quantify diffusion in small samples under load, we apply the
``deformation-diffusion'' decomposition \cite{wang13pnas,falk98pre,LiRLDL15},
\begin{equation}
\label{eq:1}
D_{i}^2=\frac{1}{N_{i}}\min_{\mathbf{J}_{i}}\sum _{j\in N_i}\mid
\mathbf{d}_{ji}^0\mathbf{J}_{i}-\mathbf{d}_{ji}\mid ^2,
\end{equation}
where $i,j$ index atoms, $D^2_i$ is a measure of magnitude of
non-affine motion of atoms around $i$; $j\in N_i$ are $i$'s initial
neighbors at the reference configuration, $\textbf{d}^0_{ji}$ is the
distance vector between atom $j$ and $i$ at the reference
configuration, and $\textbf{d}_{ji}$ is the current distance vector. The
local deformation gradient $\mathbf{J}_i$ is numerically optimized to
minimize $D^2_i$. On the right-hand side,
$\mathbf{d}_{ji}^0\mathbf{J}_{i}$ stands for the displacive deformation, while
$\mathbf{d}_{ji}^0\mathbf{J}_{i}-\mathbf{d}_{ji}$ refers to the contribution of the
non-affine, or diffusional part of the displacement \cite{wang13pnas}.
When the contacting bodies were compressed to deformed plastically, we
can see the mean $D^2$ fluctuates as that shown in Fig.3(a) due to the
dislocation plasticity and structure collapse. It is shown that $D^2$
for $R=1.5$nm contact is the largest indicating a clearly enhanced
atom diffusion.  We  borrow the threshold value from
Lindemann criterion \cite{mei07}, which was used to predict the melting
point of surface confined materials, to qualitatively compare the
extent of diffusion  during loading. For simplicity we label an atom as
diffusive when its $\sqrt{D_{i}^2}$ exceeds the $10\%$ of the nearest
neighbor distance. Fig.\ref{fig:3}(b) shows that a smaller contact
contains a higher ratio of diffusive atoms, an observation consistent
with the nanowire and nanoparticle
experiments \cite{gulseren95prb,guisbiers08,mei07}.  In our $R=2$nm
sample, the surface diffusion results in lower plastic flow stress and better adhesion. Even though
diffusion is clearly accelerated by the atomic random thermal motion
when the temperature increases \cite{sorensen96prl}, we can also
observe stress-induced surface diffusion even at $T\rightarrow0$K (see our
energy minimization simulation results in Supporting Information). It
was suggested that this type of diffusion can be not only
thermally-activated but also driven by externally applied stress \cite{LiRLDL15}
 and/or surface tension \cite{zhang05apl,Xie15WangNM}.

\begin{figure}[htp]
\centerline{\includegraphics[width=13cm]{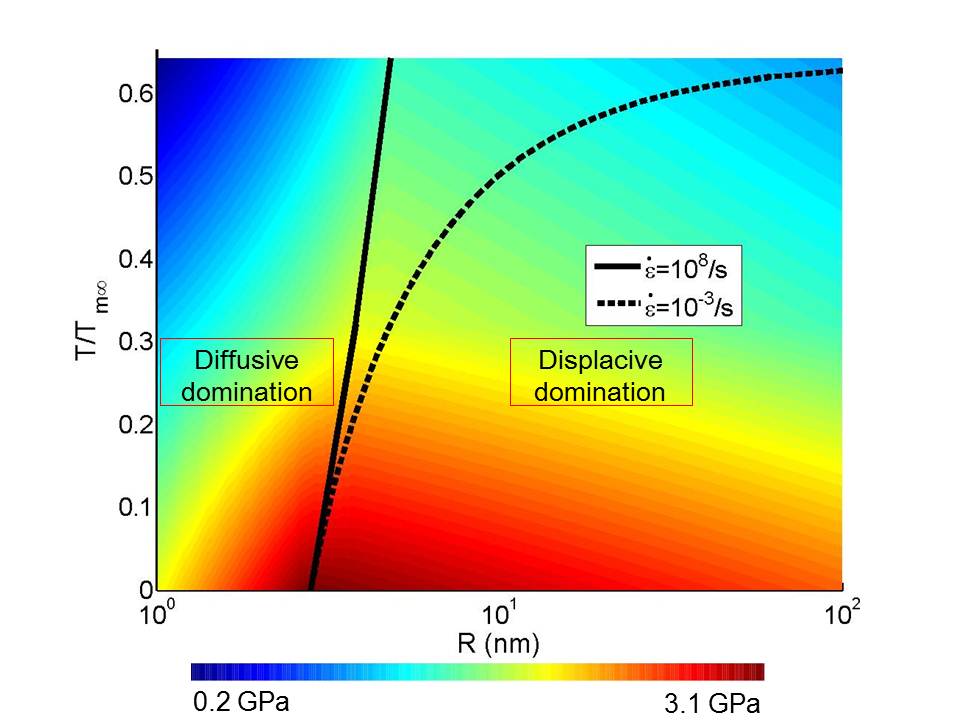}}
\caption{\label{fig:4} Contact strength mechanism map as a function of
the radius of curvature ($R$) and homologous temperature
($T/T_{m\infty}$). The color range represents the strength values. The
highest contact asperity strength computed with MD at
$\dot{\varepsilon}=10^8$/s is outlined by the solid curve. The dashed
curve represents the ideal contact strength under an ordinary
experimental strain rate.}
\end{figure}

The results above showed that the contact becomes ``smaller is
weaker'' when the surface atom diffusion dominates. Based on our
simulation data, we obtain a comprehensive contact
size-strength-temperature map in Fig.\ref{fig:4}, illustrating the
competition between the displacive and diffusion mechanisms. We find
the strongest size $R_{c}$ and homologous temperature $T/T_{\rm m\infty}$ from
simulations can be well-fitted as follows,
\begin{equation}\label{eq:S_6}
 R_{\rm c} \;=\; A \frac{T}{T_{\rm m\infty}} + B.
\end{equation}
We obtain
$A=3.1$nm and $B=2.8$nm by fitting to our MD simulation data. However, this
result should not be directly applicable to laboratory experiments
since the strain rate of the MD simulations could be many orders of
magnitude higher. To overcome this limitation, we use an empirical velocity-modified
temperature approach \cite{macgregor45} based on the Zener-Hollomon
parameter that bridges the strain rate and the temperature. This
approach considers that increasing the strain rate has the similar effect as
decreasing the temperature upon the stress-strain relation \cite{zener44}.
This semi-empirical relation bridges the temperature and strain rate as:
\begin{equation}\label{eq:S_7}
  T_{\rm exp} = T_{\rm MD} \left(1-\frac{k_{\rm B}T_{\rm exp}}{Q_{\rm
  S}}\ln\frac{\dot{\varepsilon}_{\rm MD}}{\dot{\varepsilon}_{\rm exp}} \right),
\end{equation}
where $Q_{\rm S}$ is an activation energy, $\dot{\varepsilon}_{\rm MD}$ is the simulation
strain rate, $T_{\rm exp}$ is the experimental temperature
and $\dot{\varepsilon}_{\rm exp}$ is the 
experimental strain rate. The Zener-Hollomon corrected strongest size can be derived when we combine Eqs.\ref{eq:S_6} and \ref{eq:S_7} to write

\begin{equation}\label{eq:S_8}
   R_{\rm c} \;=\;
   \frac{T_{\rm exp}}{T_{\rm m\infty}}\left(\frac{A}{1-\frac{k_{\rm B}T_{\rm exp}}{Q_{\rm
  S}}\ln(\frac{\dot{\varepsilon}_{\rm MD}}{\dot{\varepsilon}_{\rm exp}})}\right)+B.
\end{equation}
We assume that $\dot{\varepsilon}_{\rm exp}=10^{-3}$/s as a typical
laboratory experiment strain rate, and $Q_{\rm S}=126$kJ/mol = $1.3$eV for the
aluminium system studied here \cite{medina96}, which should be an
upper bound for processes controlled by surface diffusion (and therefore
gives the most sensitive strain-rate dependence). The predicted
temperature-size effects at $\dot{\varepsilon}_{\rm exp}=10^{-3}$/s is
shown by the dashed curve in Fig. \ref{fig:4}. The sub-10-nm Au tips
\cite{strachan06nano} and Ag particles \cite{sun14nat} at homologous
temperature $T_{\rm exp}/T_{\rm m\infty}$ of $0.22$ and $0.24$ are then in the
diffusion-dominated regime, which are in agreement with the
experimental observations \cite{strachan06nano,sun14nat}.

Bridging the gap between nanoscale contacts and the electrical,
thermal and mechanical properties of rough macroscopic interfaces
\cite{greenwood66,persson06,pastewka14pnas,akarapu11prl} must require
accurate information about the size-dependent plasticity. From
Fig.\ref{fig:1}, it can be seen that the strength drops precipitously\cite{tian13SR}
when the asperity size goes below $R_{\rm c}$. The plastic
deformation strength, which was
often considered a constant, 
is clearly a function of the asperity size. Moreover, the correlation
between the critical size and temperature/strain rate provides some
physical basis for $\lambda_s$, and also a criterion to judge whether
the asperity in contact is in the diffusion-controlled regime
(Fig.\ref{fig:4}), which if so is expected to bond more strongly. 
Such criterion
may be applied to material cold welding
\cite{ferguson91science,lu10nat} and self assembly
\cite{klajn07science}, 
and for physics-based modeling of the electrical, thermal
and mechanical properties of contacts.

\acknowledgement
This work is supported by a grant-in-aid of 985 Project from Xi'an
Jiaotong University, the National Natural Science Foundation of China
(Grant No. 11204228) and the National Basic Research Program of China
(2012CB619402 and 2014CB644003). JL acknowledges support by NSF
DMR-1410636 and DMR-1120901.

\suppinfo
Surface atom diffusion to contact necking region, crystal orientation combination dependence of the contact strength, Stress-strain data for explaining the strength measure. These materials are available free of charge via the Internet at http://pubs.acs.org.

\end{document}